\documentstyle[12pt]{article}
\begin{document}
\thispagestyle{empty}
\renewcommand{\thefootnote}{\dag}

\hfill UB-ECM-PF 94/11

\hfill April, 1994

\vspace*{3mm}

\begin{center}

{\large \bf Asymptotic regimes in quantum gravity at large distances \\}

{\large \bf and running Newtonian and cosmological constants \\}

\vskip 0.5truecm

\bigskip
\renewcommand{\thefootnote}{1}

{\bf E. Elizalde}
\footnote{E-mail: eli@zeta.ecm.ub.es, eli@ebubecm1.bitnet}
\\
Center of Advanced Studies, CSIC, Cam\'{\i} de Sta.
B\`arbara, 17300 Blanes, \\
and Department E.C.M., Faculty of Physics,
University of  Barcelona, \\  Diagonal 647, 08028 Barcelona,
Catalonia,
Spain \\

\renewcommand{\thefootnote}{2}

\vskip 0.5truecm
{\bf S.D. Odintsov}
\footnote{E-mail: odintsov@ebubecm1.bitnet}
\\
Department E.C.M., Faculty of Physics,
University of  Barcelona, \\  Diagonal 647, 08028 Barcelona,
Catalonia,
Spain \\
and Tomsk Pedagogical Institute, 634041 Tomsk, Russia \\

\vskip 0.25truecm
and
\vskip 0.25truecm

\renewcommand{\thefootnote}{3}

{\bf I.L. Shapiro}
\footnote{E-mail: shapiro@fusion.sci.hiroshima-u.ac.jp}
\\
Tomsk Pedagogical Institute, 634041 Tomsk, Russia \\
and Department of Physics, Hiroshima University,
Higashi-Hiroshima 724, Japan

\end{center}
\vskip 1.5truecm

\noindent
{\bf Abstract}. We consider
a multiplicatively renormalizable higher-derivative scalar theory which
is used as an effective theory for quantum gravity at large distances
(infrared phase of quantum gravity). The asymptotic regimes (in
particular, the asymptotically free infrared regime) for the
coupling constants ---specifically the
Newtonian and the cosmological constant--- are obtained.
The running of the Newton and cosmological constants in the infrared
asymptotically free regime may be relevant for solving the
cosmological constant problem and for estimating the leading-log
corrections to the static gravitational potential.

\setcounter{page}1
\renewcommand{\thefootnote}{\arabic{footnote}}
\setcounter{footnote}0
\newpage

{\bf 1. Introduction.} There is a longstanding hope that the
cosmological constant problem (for a review
see [1]) may be solved with the help of quantum gravity (QG) which, on
the other hand, does not yet exist as a consistent theory.
Recently a very interesting approach to solve the
cosmological constant problem based on an effective theory
of the conformal factor
has been suggested in Ref. [2] (see, also [3]). The effective theory for
the
conformal factor, obtained as a result of integrating over the conformal
anomaly [4] (the derivation of the anomaly induced action is done in
[7,8]),
has been used to describe QG at large distances (larger then the horizon
scale)
in the far infrared (IR) [2]. (For a different approach to QG in the far
infrared, see
[5]). The anomalous scaling dimension for the conformal factor has been
derived in the IR
stable fixed point, what gives a way to fix the cosmological
constant. Note
that the main idea of [2] is very similar to the
standard approach to the Polyakov
action in two dimensions.

In the present letter we discuss the effective theory of the conformal
factor
aiming at the description of QG at large distances from a different
viewpoint in
comparison with [2]. We shall not consider this theory exactly in the IR
stable
fixed point. Instead, we shall write the renormalization group (RG)
equations
(corresponding to the more general model of [6]) for all the
effective couplings of the theory.
The special asymptotically free solutions for the effective couplings
generalize the effective theory of the conformal factor of Ref.
[2]. We find a general solution which has two different asymptotical
regimes consistent in the IR (large distances) or in the UV limit. These
RG solutions are used to discuss the running of
the effective Newtonian and cosmological constants.
Some possible applications to the case of a static gravitational
potential are pointed out.
\medskip

{\bf 2. Solutions of the RG equations for the general model.}
Let us start from the model of Ref. [6] which is renormalizable in a
generalized sense. The classical action has the form
$$
S= \int d^4x \sqrt{-g} \{ b_1 (\varphi ) \left( \Box
\varphi \right)^2 +  b_2 (\varphi ) \left( \nabla_\mu \varphi
\right) \left( \nabla^\mu \varphi \right)
 \Box \varphi +  b_3 (\varphi ) \left[ \left( \nabla_\mu \varphi
\right) \left( \nabla^\mu \varphi \right) \right]^2 +
$$
$$
 +  b_4 (\varphi ) \left( \nabla_\mu \varphi \right) \left(
\nabla^\mu \varphi \right)+  b_5 (\varphi ) +  c_1 (\varphi ) R
\left( \nabla_\mu \varphi \right) \left( \nabla^\mu \varphi
\right) +  c_2 (\varphi ) R^{\mu\nu} \left( \nabla_\mu \varphi
\right) \left( \nabla_\nu \varphi \right)
$$
$$
 + c_3 (\varphi ) R \Box \varphi + a_1 (\varphi )
R^2_{\mu\nu\alpha\beta }  + a_2 (\varphi ) R^2_{\mu\nu} + a_3
(\varphi ) R^2 + a_4 (\varphi ) R \}.
\eqno(1)
$$

Note that $[\varphi] = 0$ and the only $\varphi$ is a quantum field,
that is, the metric is purely classical.
We suppose that $c_{1,2}, a_{1,2,3,4}, b_{1,2,3}$ are constants and that
$c_3(\varphi) = c_{31}\varphi + c_{32}$. Then this theory is
multiplicatively renormalizable in the usual sense (for the above
choice of $b_4(\varphi )$, $b_5(\varphi )$).
The renormalization group equations for all the effective couplings
(supposing the standard renormalization of the scalar $\varphi$) have
been derived in [6]
$$
\frac{db_2(t)}{dt} = -10b_2(t)b_3(t) + \frac{15}{4} b_2^3(t),
b_2(0)=b_2,
$$
$$
\frac{db_3(t)}{dt} = -10b_3^2(t) +5 b_2^2(t)b_3(t),
b_3(0)=b_3,
\eqno(2)
$$
$$
\frac{dc_{31}(t)}{dt} = \frac{5}{2} b_2^2(t)c_{31}(t),
c_{31}(0)=c_{31}.
\eqno(3)
$$
$$
\frac{dc_1(t)}{dt} = \frac{2}{3} b_3(t) \left[ 9c_{31}(t)-9c_1(t)
-2c_2(t) +2 \right]+ \frac{5}{2} b_2^2(t)c_1(t), c_1(0)=c_1,
$$
$$
\frac{dc_2(t)}{dt} =-\frac{1}{3} \left[ 2c_2(t)b_3(t)-b_2^2(t)
+4b_3(t) \right]+ \frac{5}{2} b_2^2(t)c_2(t),  c_2(0)=c_2,
$$
$$
\frac{dc_{32}(t)}{dt} = \frac{1}{3}b_2(t) \left[
-9c_1(t) -2c_2(t) +2 \right]+ \frac{5}{4}
b_2^2(t)c_{32}(t),  c_{32}(0)=c_{32}.
\eqno(4)
$$
Finally, the RG equations for the effective vacuum couplings are
$$
\frac{da_1(t)}{dt} =0,  a_1(0) = a_1,
$$
$$
\frac{da_2(t)}{dt} =-\frac{1}{12}c_2^2(t) - 2 c_2(t) -
\frac{1}{15},  a_2(0) = a_2,
$$
$$
\frac{da_3(t)}{dt} =\frac{-3[4c_1(t)+c_2(t)-4c_{31}(t)]o2 +
c_2^2(t)}{48} - \frac{2c_{31}(t) -2c_1(t)-c_2(t)}{6} -
\frac{1}{30},  a_3(0) = a_3,
$$
$$
\frac{da_4(t)}{dt} =\frac{1}{2}b_4(t) \left[4c_{31}(t)
-4c_1(t)-c_2(t) + \frac{2}{3} \right],  a_4(0) = a_4.
\eqno(5)
$$
Below we consider the general and special asymptotic
regimes for the renormalization group equations
(2) and then concentrate our attention on the following particular
choice of functions $b_4 (\varphi)$ and $b_5 (\varphi)$
$$
b_4 (\varphi) = \gamma \exp ( 2\alpha \varphi),
b_5 (\varphi) = {\lambda} \exp (4 \alpha \varphi),
\eqno(6)
$$
which corresponds to the model of Ref. [2].

The general solution of equations (2) possesses some very interesting
features.
For the sake of simplicity we introduce new variables $x,y,z$ according to
$4z = b_2^2, y = b_3, x = z/y$. Then one can easily obtain the
equation $$
\frac{dy}{y} = \frac{(2x - 1)dx}{x^2 - x} \eqno(7)
$$
and, moreover, three special solutions for (2):
$$
z = y, \;\;\; \mbox{that  is} \;\;\; b_2^2 = 4b_3, \eqno(8a)
$$
$$
 z = 0, \;\;\; \mbox{that is} \;\;\;  b_2^2 =  0,     \eqno(8b)
$$
$$
y = 0, \;\;\; \mbox{that is} \;\;\;  b_3 = 0.       \eqno(8c)
$$
Let us now explore (7). This equation leads to the following relation
between $y$ and $z$
$$
z = y \left[\frac{1}{2} \pm \sqrt{\frac{y}{c_1} + \frac{1}{4}}\right].
\eqno(9) $$
Substituting (9) into the first Eq. of (2), we get
$$
\frac{dy}{y^2 \sqrt{1 + cy}} = \pm 10 dt, \ \  \;\;\;
c = \frac{1}{4c_1}. \eqno(10a)
$$
Integrating (10), we finally obtain
$$
\pm 10 t + c_2 = \frac{1}{2} \ln\left|\frac{v-1}{v+1}\right| -
\frac{1}{2}\frac{v}{v^2-1}, \hspace{2cm}
v^2 = 1 + cy.  \eqno(10b)
$$
Eq. (9) can be viewed as a first integral of the system (2), in fact it
can be rewriten in the form:
$$
b_2^4 -4 b_2^2 b_3 =k_1 b_3^3,
$$
which is a family of curves on the plane $b_2, \ b_3$ with parameter
$k_1$ (constant of motion) to be determined from the initial values
$b_2(0)$ and
$b_3(0)$. Another first integral can be easily obtained by reducing the
system (2) to an exact diferential equation by means of the integrating
factor $b_2^9b_3^8$. We then get
$$
12 b_2^{10} b_3^{10} - 5 b_2^{12}b_3^9 = k_2.
$$
Again, the parameter $k_2$ (a second constant of motion) is to be fixed
from the initial values $b_2(0)$ and $b_3(0)$.
It can be seen immediately that, in general, the two families of curves
do {\it not} intersect (whenever they do they yield a trivial, constant
trajectory). From the last expression we get the following
particular solution: if $b_3(0) = (5/12) b_2(0)^2$ then, for any $t$,
$b_3(t) = (5/12) b_2(t)^2$ and solving for the dynamics (2) one obtains
$$
b_2(t) = \frac{b_2}{\sqrt{1+ \frac{5}{6} b_2^2 t}}, \ \ \ \
b_3(t) = \frac{5b_2^2}{2(6+5b_2^2t)}.
$$

We observe that (9) and (10) admit two possible asymptotic
regimes for
$y$ and $z$. In both regimes $y$ tends to zero when $t\rightarrow\infty$.
The first regime corresponds to the positive sign in
(9). Then, it is easy to see that in the $t\rightarrow-\infty$ limit
(large distance) the couplings
have infinitesimal difference with the special solution (8a), where we
have $$
b_2^2(t) = \frac{2b_2^2}{2-5 b_2^2t}, \ \ \ \ \
b_3(t) = \frac{b_2^2}{2 \left( 2- 5 b_2^2t \right)}. \eqno(11)
$$
At the same time this regime leads to the zero-charge problem in the UV
limit $t\rightarrow+\infty$.

The second regime corresponds to the negative sign in (9). After tiny
algebra
we obtain that in this case $z\sim y^2$ and moreover, from (10) we get
that $y(t) \sim t^{-1}$ in both limits. So we see that this regime
is in some sense similar to the special solution (8b). Substituting
$b_2^2 \ll b_3$
into (2), we get the following regime, which can be asymptotically free
in both the  UV and IR limits [6]
$$
b_2 (t) \simeq \frac{b_2}{1 + 10 b_3 t}, \ \ \ \
b_3 (t) \simeq \frac{b_3}{1 + 10 b_3 t}. \eqno(12)
$$
Here, we have asymptotic freedom in the UV or IR limits, and the result
depends on the sign of $b_3$. It is very interesting to note that
the sign
of $b_3$ is not connected with any physical quantity (contrary to
what happens in the $\lambda
\varphi^4$ theory where the wrong sign of $\lambda$
corresponds to an unstable
classical potential). Hence one can choose any sign of $b_3$ and have
asymptotic freedom in the IR or in the UV region. (Notice, however, that
in the model [2] $b_3$ is always negative.) Therefore, the
theory under discussion posesses some unusual property in this point.
Observe also that from the discussion above it follows that one can
easily
find the running couplings corresponding to the rest of the parameters
of the theory.
\medskip

{\bf 3.  Running couplings in QG at large distances.} Now we shall study
the running of the coupling constants in the effective theory of
the conformal factor.
This theory has been discussed in Ref. [2], but only in the fixed point
$b_2 = b_3 = 0.$  On the countrary, we shall here consider all the
possible
asymptotic regimes for these couplings. Notice that the correspondence
between our model and the model of [2] is obtained by setting
$$
b_1 = - \frac{Q^2}{(4\pi)^2},\;\;\;b_2 = - 2\zeta\alpha,\;\;\;
b_3 = - \zeta^2 \alpha,\;\;\;b_4 = \gamma \exp(2\alpha\varphi),
\;\;\;b_5 =-\frac{\lambda}{\alpha^2} \exp(4\alpha\varphi), \eqno(13)
$$
where $Q^2$ is the four-dimentional central charge, $\alpha$  the
scaling dimension for $\varphi$,
 $\gamma$ the gravitational coupling,
and $\lambda$ is
the cosmological constant. Further, we will use the notations (1).
As one can see, the theory under consideration differs from the general
model (1) by the choice (6) for the couplings.
Such a theory can be renormalized in two different ways.
One can consider $b_1, b_2, b_3$ as arbitrary constants, which
have to be renormalized as well as $a_1,a_2,a_3$ and $c_1,c_2,c_3$.
Or one can either consider $b_1$ as a non-essential constant
 and renormalize $\alpha$
instead of $b_1$. Here we will adopt the second point of view.
The RG equations
 for $ b_2(t), b_3(t), c_1(t), c_2(t), c_3(t), a_1(t), a_2(t), a_3(t)$
have the same form (2)--(5). At the same time, we easily find the
RG equations for $\gamma(t),
\alpha(t), \lambda(t)$, which are a particular case of (3):
$$
\frac{d\alpha(t)}{dt} = \frac{5}{4} b_2^2(t),
\alpha(0)=\alpha, \eqno(14)
$$
$$
\frac{d\gamma(t)}{dt} = \gamma(t)[b_2^2(t) + 4\alpha^2(t) +
6b_2(t)\alpha(t)]
,\;\; \gamma(0)=\gamma, \eqno(15)
$$
$$
\frac{d\lambda(t)}{dt} = 16 \alpha^2(t) \lambda(t) - \gamma^2(t)
,\;\; \lambda(0)=\lambda.   \eqno(16)
$$

These equations can be solved indepently of (4) and (5), because
the corresponding running couplings do not change when we restrict the
theory  to live in
flat space. Next, since in all asymptotic regimes the basic couplings
$b_2,b_3$ have just the same qualitative asymptotic behaviour, we can
restrict ourselves by just one of them. Let us consider the special
solution (8b), and put $\frac{1}{4}b_3(t) = b_2^2(t) = {b_2^2}
{\left( 1-\frac{5}{2} b_2^2t \right)}^{-1}$. Then we easily find the
solution of (13) in the form
$$
\alpha(t) = \frac{\alpha}{\sqrt{1-\frac{5}{2} b_2^2t}}, \eqno(17)
$$
$$
\gamma(t) = \gamma {\left( 1-\frac{5}{2} b_2^2t
\right)}^{- \textstyle\frac{ 2 (b_2^2 + 4\alpha^2 +
6b_2\alpha)}{5b_2^2}}, \eqno(18)
$$
$$
\lambda(t) = \frac{2\gamma}{3(b_2^2 + 8\alpha^2 - 4b_2\alpha)}
{\left( 1-\frac{5}{2} b_2^2t \right)}^
{\textstyle\frac{ 3b_2^2 - 8\alpha^2 - 12b_2\alpha)}{5b_2^2}}
$$
$$
+ {\left[ \lambda - \frac{2\gamma}{3(b_2^2 + 8\alpha^2 - 4b_2\alpha)}
\right]} {\left( 1-\frac{5}{2} b_2^2t \right)}^
{- \textstyle\frac{ 32\alpha^2}{5b_2^2}}
\eqno(19) $$
That solution describes qualitatively the IR phase of quantum gravity in
frames of our effective theory. Let us now give a brief analysis of the
behaviour of the effective couplings
$\lambda(t),\gamma(t)$. The asymptotics essentially depend on the value
of the ratio $u = \alpha /b_2$. For example, if
$- 3 - \sqrt{5} < u < - 3 + \sqrt{5}$, then
$\lambda(t)\rightarrow 0$ and $\gamma(t)\rightarrow \infty$
in the IR limit. When
$u = - 3 \pm \sqrt{5}$, then $\lambda(t)\rightarrow 0$ and
$\gamma(t) = \gamma =$ const.

Let us discuss the behaviour of the running Newton and cosmological
constants in more detail. The standard expression for the Newtonian
potential has the form
$$
V(r) = - \frac{Gm_1m_2}{r}, \eqno(20)
$$
where in our notation $\gamma = 3/(8\pi G)$. Observe that, in
$R^2$-gravity, already at the classical level there appear
corrections of Yukawa type to the classical gravitational
 potential (20) (see Ref. [9], and for a review of quantum
$R^2$-gravity see [10]).

We are working in terms of an effective theory of quantum gravity.
Such a description is valid in the far infrared (large distances) and
the extremely small exponentially suppressing corrections of Ref. [9] to
the potential (20) are not important in such case. However, the fact
that the Newton coupling becomes the running coupling is very important
for its possible cosmological applications. Indeed, we are guided by the
well-known example of the electrostatic potential in quantum
electrodynamics. Here, this potential can be alternatively obtained from
the classical part of the potential by the change of electric charge to
corresponding running coupling with $t= (1/32\pi^2) \ln (r_0^2/r^2)$.
Hence, the Wilsonian (or renormalization-group improved) electrostatic
potential coincides with the explicit calculation of the
electrostatic potential in perturbation theory.

In the same way, in Refs. [11] and [12] it was suggested to use the
Wilsonian gravitational potential (20) (i.e., with running gravitational
coupling constant) for a possible solution of the dark matter problem
[12] and of the cosmological constant problem (in the wormhole context).
However, in Ref. [12] the
running Newton constant that was obtained from quantum $R^2$-gravity in
the ultraviolet asymptotically free regime was used. Surely, this is
not quite consistent for estimating the infrared effects.

Instead, we have used the running Newton coupling constant calculated in
the effective theory, with the aim to describe quantum gravity in the
far infrared (large distances). Then the leading quantum corrections for
the Newtonian potential together with the classical part yields (see
Eqs. (18) and (20)):
$$
V(r) = - \frac{G_0 m_1m_2}{r}
\left( 1-\frac{5b_2^2}{64\pi^2} \ln \frac{r_0^2}{r^2}
\right)^{- \textstyle\frac{ 2 (b_2^2 + 4\alpha^2 +
6b_2\alpha)}{5b_2^2}} $$
$$
\simeq  - \frac{G_0m_1m_2}{r}
\left( 1+
 \frac{2 (b_2^2 + 4\alpha^2 + 6b_2\alpha)}{32\pi^2} \,
 \ln \frac{r_0^2}{r^2} \right).
 \eqno(21)
$$
Here $G_0$ is the initial value of the Newton constant at distance
$r_0$. Eq. (21) shows the leading-log corrections to the classical
gravitational potential in the context of our effective theory of
quantum gravity. For a rough estimation of the quantum correction in
(21) we take $\alpha =1$ (classical value) and use $b_2= -2(2 \tilde{b}
+ 2 \tilde{b}' + 3 \tilde{b}'')$, where the conformal anomaly $T=
\tilde{b} (F + \frac{2}{3} \Box R) + \tilde{b}' G + \tilde{b}'' \Box R$
coefficients are known [4]. For instance, for scalars they are
proportional
to the effective number of matter fields. Hence, the quantum correction
to the Newtonian gravitational potential is proportional to the square
of the effective number of matter fields.  Of course, this
correction is far too
small to be experimentally measured. However, this potential (21) might
growe within the region $u< -3 - \sqrt{5}$ or $u>-3+\sqrt{5}$, and
this might lead to more reasonable implications for the solution of the
dark matter problem than just use
of the UF asymptotically free running couplings of quantum
$R^2$-gravity. Let us recall that, recently, leading corrections to
the
Newtonian potential have been calculated in the one-loop approach for
the Einstein theory considered as an effective theory of quantum gravity
[13], where logarithmic corrections were not found. In our approach we
get a definitely better approximation to the Newton
potential, because we use a Wilsonian potential that sums all leading
logarithms of the perturbation theory. This is, of course, a richer
situation in comparison with what  one gets when taking into account the
one-loop corrections only.

Notice that the running cosmological constant (19) may quickly decay in
the infrared region, as it follows from the explicit expression (see
also [14]). This
fact depends very much on the choice of the parameters of the theory
and gives an indication of the possibility of solving the cosmological
constant problem as a result of infrared effects in quantum gravity.

Concerning the behaviour of $\gamma(t) and \lambda(t)$ on the
general solutions of (2), since the solutions (11) and (12) are related
with each other by the obvious change of $t$, it is clear that the above
picture will preserve its structure when the asymptotic regime (11)
is used.
That is why it is not necessary to study the details of the solutions
for the couplings $c_{1,2,3}(t)$ and $a_{1,2,3}$. This analysis has
been already performed in a previous paper, Ref. [6], in the framework
of the asymptotic regime (12),
and will be surely  the same for the special solution (11), that is
just the starting model in Ref. [2]. Hence, we see again the
asymptotical conformal invariance of the model (1) in the infrared
asymptotic regime.

Summing up, we have considered in this work the model of
Ref. [2],  not just in the
IR stable fixed point (as in [2]), but in the most general
asymptotic regime. The theory
under consideration possesses some unusual properties. For instance,
there are in the
theory some asymptotic regimes which are free of the zero-charge
problem, either in the UV or in the IR limit. We have then considered
the RG equations
for the Newtonian and for the cosmological constants, and found that in
both the UV and IR
regions the theory may exhibit
asymptotic regimes with decreasing cosmological
constant and (or) increasing Newton running coupling. An estimation of
the
leading-log corrections to the static gravitational potential has been
also given.
It would be of interest to construct the supersymmetric version of the
theory (1) (for a general introduction to supersymmetric theories, see
[15]) which should describe the superconformal anomaly induced action.
Then, using a technique similar to the one above we would be able to
provide a description of quantum supergravity at large distances.
\vspace{5mm}

\noindent{\large \bf Acknowledgments}

The authors thank R. Tarrach for helpful remarks.
SDO and ILS would like to acknowledge the kind hospitality of the
members of the Dept. ECM, Barcelona University and of the Particle
Physics Group, Hiroshima University, respectively. This work has been
supported in part by DGICYT
(Spain), project no. PB90-0022, by CIRIT (Generalitat de Catalunya), and
by the Russian Foundation for Fundamental Research, project
no. 94-02-03234.

\end{document}